\documentclass[useAMS]{mn2e}
\usepackage{psfig}

\newif\ifAMStwofonts
\AMStwofontstrue

\def\etal{{et al.}}
\def\xte{{\it RXTE}}
\def\asca{{\it ASCA}}
\def\xmm{{\it XMM-Newton}}
\def\bhm{{M$_{\rm BH}$}}
\def\rms{{$\sigma^{2}_{\rm nxs}$}}
\def\bhml{{M$_{\rm BH, L}$}}
\def\bhms{{M$_{{\rm BH}, \sigma_{\rm nxs}^2}$}}
\def\nb{{$\nu_{\rm bf}$}}

\def\psdamp{{$PSD_{\rm amp}$}}
\def\lx{{$L_{\rm X}$}}
\def\T{{$T$}}

\title[Black hole masses in AGN]
  {Black hole mass estimation from X-ray variability measurements in AGN}
\author[M. Nikolajuk \etal]
  {M.~Nikolajuk$^{1}$, I.E.~Papadakis$^{2}$ and B.~Czerny$^{1}$ \\
  $^1$ N. Copernicus Astronomical Centre, Bartycka 18, 00-716 Warsaw, 
  Poland \\
  $^2$ Physics Department, University of Crete, 71 003, Heraklion, Crete,
  Greece }

\def\LaTeX{L\kern-.36em\raise.3ex\hbox{a}\kern-.15em
    T\kern-.1667em\lower.7ex\hbox{E}\kern-.125emX}

\begin{document}

\label{firstpage}

\maketitle

\begin{abstract} 
We propose a new method of estimation of the black hole masses in AGN
based on the normalized excess variance, \rms. We derive a relation
between \rms, the length of the observation, \T, the light curve bin
size, $\Delta t$, and the black hole mass, assuming that (i) the power
spectrum above the high frequency break, \nb, has a slope of $-2$, (ii)
the high frequency break scales with black hole mass, (iii) the power
spectrum amplitude (in $frequency \times power$ space) is universal
and (iv) \rms\ is calculated from observations of length $T < 1/$\nb.  
Values of black hole masses in AGN obtained with this method are
consistent with estimates based on other techniques such as
reverberation mapping or the \bhm-stellar velocity dispersion relation.
The method is formally equivalent to methods based on power spectrum
scaling with mass but the use of \rms\ has the big advantage of being
applicable to relatively low quality data.
\end{abstract}


\section{Introduction}

The X-ray emission of active galactic nuclei (AGN) displays variations
over a wide range of time scales.  The first convincing demonstration
of this phenomenon came with the {\it EXOSAT} `long looks' (Lawrence
\etal\ 1987; McHardy \& Czerny 1987). The data showed no
characteristic time-scales, and the power spectral density function
(PSD) showed a `red-noise' shape. At the same time, it was also
noticed that more luminous sources show slower variations. Various
methods have been used in the past in order to determine the X-ray
variability amplitude of AGN: determination of the `two-folding'
time-scale (i.e. the time-scale for the emitted flux to change by a
factor of two), calculation of the PSD amplitude at a given frequency,
and estimation of the so called `normalized excess variance' (\rms,
i.e.  the variance of a light curve normalized by its mean squared
after correcting for the experimental noise). In all cases, these
quantities appear to anti-correlate with the source luminosity (Barr
\& Mushotzky 1986; Lawrence \& Papadakis 1993; Green, McHardy \&
Lehto 1993; Nandra \etal\ 1997; Turner \etal\ 1999; Leighly 1999;
Markowitz \& Edelson 2001).  One possible explanation for the observed
anti-correlation between the X-ray variability amplitude and X-ray
luminosity, \lx, is that more luminous AGN have larger black hole
masses (\bhm) as well.  In this case, as \bhm\ increases, so does the
size of the X-ray source, and a change of the source luminosity by a
constant fraction takes relatively longer; equivalently, variability
amplitude measured at a fixed timescale should decrease with mass. The
significant progress in the measurement of the black hole (BH) mass in
the centers of nearby galaxies has made it possible to actually
directly test the dependence of the X-ray variability amplitude on
\bhm\ in AGN. The data show that \rms\ measured at a certain timescale
is indeed anti-correlated with \bhm\ (Lu \& Yu 2001; Bian \& Zao 2003;
Papadakis 2004).

If there is an intrinsic correlation between X-ray variability and
\bhm, then X-ray variability measurements could be used in order to
measure the central black hole mass in these objects.  This
possibility has already been studied in the past. Hayashida \etal\
(1998) and Czerny \etal\ (2001) have used the PSD normalized by the
square of the mean flux as a measure of the X-ray variability of
a source. By calculating the ratio of the frequencies at which the
PSD$\times frequency$ has a certain value in AGN and Cyg X-1, they
were able to estimate the \bhm\ in AGN.  Recently, long, well-sampled
\xte\ light curves have been used in order to accurately estimate the
X-ray PSD of AGN. One of the main results from these studies has been
the unambiguous detection of a characteristic `break frequency', \nb,
at which the power spectrum changes its slope from a value of $-2$ to
$-1$ above and below \nb, respectively (Uttley \etal\ 2002; Markowitz
\etal\ 2003). This break frequency is analogous to the high frequency
break in galactic sources and should not be confused with the low
frequency break, where the spectrum changes slope from $-1$ above to
$0$ below.  This high frequency break appears to correlate well with
\bhm, in the sense that $1/\nu_{\rm bf} \propto {\rm M_{BH}}$
(Markowitz \etal\ 2003). Thus another way to estimate \bhm\ is through
the determination of \nb\ in the X-ray PSD of a source. However, this
method requires rather long, high quality data.

We propose a new method to estimate \bhm\ in AGN, which may be used even
for relatively low quality and/or short data coverage. The method is based
on the use of \rms\ as a measure of their X-ray variability amplitude, and
takes Cyg X-1 as a reference point. Using recent, archival \xte\ and
\asca\ light curves of 21 AGN, we estimate their \bhm, and we compare our
results with \bhm\ estimates which are based on other methods.  We
find that the accuracy of the \bhm\ estimates using our new method is
roughly similar to the accuracy of the estimates with well established
methods like reverberation mapping. Furthermore, the \rms\ estimation
of an X-ray light curve is rather easy and straightforward.
Consequently, as long as typical AGN observations performed by present
day satellites (like {\it XMM-Newton} and {\it CHANDRA}) detect
significant variations, they can be used to estimate \rms\ and hence
\bhm, using the method which we present in this work. It is possible
then that the use of our method can provide \bhm\ estimates for a
large number of AGN in the near future.

\section{The relation between BH mass and excess variance}   

The scaling of the PSD with mass is most conveniently discussed after
multiplying the normalized PSD by the frequency $\nu$. We assume here that 
the break
frequency, \nb, scales with the mass: $\nu_{\rm bf} \propto {\rm
M_{BH}}^{-1}$, but the amplitude of the normalized PSD $\times \nu$
part at the frequency break does not depend on mass. We
specifically define this normalization at \nb\ and call it the PSD
amplitude,

\begin{equation}
PSD_{\rm amp} = P(\nu_{\rm bf}) \times \nu_{\rm bf},
\end{equation}
where $P(\nu)$ is the power spectrum normalized to the mean squared.  
Recently, Papadakis (2004) has showed that this quantity is indeed similar
in a few well studied Seyfert 1 galaxies.

Let us suppose that an AGN is observed for a period $T<1/$\nb, 
and that the resulting light curve is binned using a bin size of
$\Delta t$.  The excess variance of the light curve is equal to the
integral of the PSD,

\begin{equation}
\sigma^{2}_{\rm nxs}=\int_{1/T}^{1/2\Delta t}P(\nu)d\nu 
\end{equation}
Let us also suppose that $P(\nu)=A(\nu/$\nb$)^{-2}$, where
$A=$\psdamp/\nb ~ and \psdamp\ is constant, and that \nb\ = $B$/\bhm,
where $B$ is another constant. Under these assumptions,

\begin{equation}
\sigma^{2}_{\rm nxs} = PSD_{\rm amp} \frac{B}{{\rm M}_{\rm 
BH}}(T-2\Delta t).
\end{equation}
The equation above can be re-written as follows,
\begin{equation}
{\rm M}_{\rm BH} = C(T-2\Delta t)/\sigma^{2}_{\rm nxs},
\label{bhm}
\end{equation}
where $C=$ \psdamp $B$. Eq.~(\ref{bhm}) can be used to estimate
\bhm\ provided the \rms\ of a source has been estimated, and $C$ is
known. In order to determine $C$, we used the results from the PSD
analysis of a Galactic black hole candidate, namely Cyg X-1.

The X-ray PSD of this source has been extensively studied the last twenty
years or so (Kemp et al. 1978; Gierli\'nski et al. 1999; Nowak et al.  
1999; Churazov et al. 2001; Pottschmidt et al. 2003). In its low/hard
state, the PSD of the source can be roughly represented as a power law
with a high frequency slope of $-2$ , which breaks to a slope of $-1$ below
$2-3$ Hz, and then to a slope of zero below $\sim 0.2-0.02$ Hz (Nowak et
al. 1999; Pottschmidt 2003).  Assuming that Eq.~(\ref{bhm}) holds for Cyg
X-1 as well, and adopting the value of 10 M$_{\odot}$ for its black hole
mass (Nowak et al. 1999; Gierli\'nski et al. 1999 and references therein),
we determined the value of $C$ as follows.

We used the results from the PSD analysis of 130 \emph{RXTE}
observations of Cyg X-1 reported by Pottschmidt et al. 2003. They
calculated the PSD for each observation and fitted them using
Lorentzian functions. The RXTE observations cover a long period of
time (from December 1997 to July 2001) in which Cyg X-1 was mainly in
its hard state.  We chose 68 observations, carefully avoiding data
from soft or failed transition states. We calculated the excess variance
using the relation \rms=$\int_{10 Hz}^{+\infty} P(\nu) d\nu$, where
the lower frequency limit was chosen in such a way that it will always
be higher than \nb\ in this source.  Using Eq.~(4), with
\bhm=10 M$_{\odot}$, $\Delta t=0$, and $T=1/(10$ Hz)$=0.1$ sec, we
find that $C$ varies between $\sim 0.7$ and $\sim 1.3$, with the
arithmetic mean being $C=0.96\pm0.02$. We accept this value for the
normalization constant $C$ in Eq. (\ref{bhm}).

\section{New \bhm\ estimates for 21 AGN}

In order to investigate the accuracy of our method, we selected for
analysis several AGN with known black hole masses and recently
obtained \xte\ or \asca\ light curves. First, we estimated the \rms\
of each light curve using the relation (Nandra et al., 1997),

\begin{equation}
\sigma^{2}_{\rm nxs}= \frac{1}{N_{\rm data}\,\bar{x}^{2}} 
\sum_{i=1}^{N_{\rm data}} [(x_{i}-\bar{x})^{2}-\sigma_{i}^{2}] \,,
\label{rms}
\end{equation}
where $N_{\rm data}$ is the number of data points, and $\bar{x},
\sigma_{i}$ are the unweighted, arithmetic mean and error of the
$x_{i}$, respectively. Then, using Eq. (\ref{bhm}) we estimated
\bhm, and compared our \bhm\ estimates with the existing values from
literature based on the reverberation mapping technique or the
measurement of stellar velocity dispersion (Section~\ref{res},
Table~1).

\begin{table}
\label{tab.mbh}
\caption{Previous and new \bhm\ estimates for 21 AGN}
\begin{tabular}{@{}lccl}
\hline
Name & \bhml & \bhms & Proposal \\
 & ($\times 10^{7}$M$_{\odot})$  & ($\times 10^{7}$M$_{\odot})$ & number 
\\
\hline
3C120     & $6.3$  (1,2)& $14.5$  & R/P30404(3) \\
3C390.3   & $35.5$ (1)  & $26.8$  & R/P10340(7) \\
Akn120    & $18.6$ (1)  & $5.6$   & R/P30232(1) \\
IC4329A   & $0.6$  (1)  & $12.3$  & R/P40153(2) \\
 && &R/P20315(2)\\
 && &R/P10313(4)\\
 && &R/P50706(1)\\
IC5063 (S2)&$5.5$  (2)  & $(0.19)$& R/P10337(1) \\
 &&3.8 &\\
Mrk348 (S2)&$1.6$  (2)  & $7.5$ & R/P10326(3) \\
Mrk509    & $7.5$  (1)  & $7.0$ & R/P10311(8) \\
Mrk766 ($\star$)  & $0.35$ (2)& $(0.0066)$ & R/P60135(23) \\
 &&0.132&\\
NGC3227   & $4.1$  (1,3)& $2.2$ & R/P40151(6) \\
 && &R/P10292(3)\\
NGC3516   & $1.7$  (3)  & $1.7$ & R/P50159(1) \\
 && &R/P20316(4)\\
 && &R/P30224(2)\\
NGC3783   & $1.0$  (1)  & $0.83$& R/P10297(2,1) \\
 && &R/P30227(1)\\
NGC4051 ($\star$) & $0.05$ (5)& $(0.0014)$ & R/P50153(128) \\
 &&0.028&\\
NGC4151   & $1.4$  (1)  & $1.9$ & R/P50155(2) \\
 && &R/P00022(3)\\
 && &R/P00024(4)\\
NGC4395   & $0.007$(6)  & $0.022$& A/76006000(2) \\
NGC4593   & $0.66$ (3)  & $0.60$ & A/71024000(3) \\
 && &A/75023010(1)\\
NGC5506 ($\star$) & $8.8$ (4)& $(0.32)$ & R/P20318(10) \\
 &&6.4&\\
NGC5548   & $9.2$ (1,7) & $12.3$  & R/P30220(2) \\
 && &R/P10297(3)\\
 && &R/P30218(3)\\
NGC7469   & $0.7$ (1)   & $1.3$ & R/P10315(20) \\
PG0052+251 & $26.1$(1)& $10.0$ & R/P40157(1) \\
 && &R/P20338(1)\\
PG0804+761 & $7.6$ (1)& $5.6$  & R/P40157(3) \\
PG1211+143 ($\star$) & $3.2$ (1)& $(0.40)$ & A/70025000(1) \\
 &&8.1&\\
\hline
\end{tabular}
Col.~(1) lists the object name and type: (S2) denote Seyfert 2 objects,
($\star$) - Narrow Line Seyfert 1. All other objects are classified as
Seyfert 1 or quasars. Col. (2) lists the \bhml\ values taken from
literature. The numbers in parentheses in Col.~(2) correspond to the
following references: (1) Kaspi \etal\ (2000), (2) Woo \& Urry (2002), (3)
Onken \etal\ (2003), (4) Papadakis (2004), (5) Shemmer \etal\ (2003), (6)
Filippenko \& Ho (2003), (7) Peterson \& Wandel (2000).  Col.~(3) lists
the \bhms\ values obtained in this work. For NLS1s, the values in
parentheses are the \bhm\ values found using Eq. (3), while the
values listed below are multiplied by a factor of 20 (see text for
details). Col.~(4) lists the observation details. The first letter refers to
the satellite (R -- RXTE, A -- ASCA), the number that follows refers to
proposal number, and we list the number of light curve parts,
$N_{\rm p}$, which we used in order to determine \rms, in parentheses.
\end{table}

\subsection{Remarks about preparation of the data}
\label{prescr}

\xte\ has regularly observed a large number of AGN over the last few 
years, providing long light curves with good signal to noise
(S/N). For almost all the objects listed in Table~1 we have retrieved
$2-9$ keV PCA background subtracted light curves from the standard
data products (`StdProds') archive of \xte\ (detailed information
about the data in this archive is given at the \xte\ mission web
site). There are no archival \xte\ data for 3 AGN with previously
estimated \bhm\ (namely PG1211+143, NGC4395 and NGC4593). For these
sources we retrieved $2-10$ keV, ASCA light curves from the TARTARUS
database. We used background subtracted GIS and SIS light curves which
we combined in all cases, except from the first observation of
NGC4593. In this case we used the GIS data only, as there were many
missing points in the SIS light curve.

The character of the obtained light curves depended mostly on the type
of the monitoring. A number of the light curves came from sparse
extensive monitoring. The source was observed typically for 10~ksec
once per day, for several days. For example, 3C390.3 was observed
usually once per day, for 2 months, with a single observation lasting
about 3~ksec. Such light curves were binned to 5400 sec (roughly the
orbital period of the satellite). If the resulting light curve
consisted predominantly of single points separated by a day, then the
remaining few consecutive data points were averaged to create a
uniformly covered light curve.

Other light curves came from intensive monitoring. The source was
observed continuously for 40 ksec or more, with gaps only caused by
earth occultation. For example, NGC3516 was observed for over one
day.  The observational campaign was sometimes repeated or supplemented by
sparse extensive monitoring. NGC7469 was almost continuously observed
for about 1 month: 10 to 20~ksec exposures were made almost every day.
Such light curves were also binned to 5400~s.

Several sources were occasionally observed for $10-40$~ksec. The bin
size used for these light curves typically varied from 512~s to
5400~s, depending on the length of the data and the S/N ratio.

The light curve of PG0052+251 was binned to 10800~s in order to
increase the S/N ratio. The light curve from continuous monitoring of
NGC4593 was binned to 2048~s and divided into separate sets of 30~ksec
duration in order to estimate \rms\ in more than 2 sub-parts.  Even
shorter bin size was requested for NLS1 sources, as discussed
below.

The long light curves were divided into separate shorter strings of
length \T, as required by the condition $T<1/$\nb\ (see Section 2). The
value of \nb\ for each source was estimated from the previous \bhm\
estimates, using the 1/\nb-\bhm\ relation of Markowitz et al.~(2003). The
PSD break frequencies of 3~NLS1 objects (NGC4051, NGC4253/Mrk766, and
NGC5506) are high (see McHardy et al. 2003; Vaughan \& Fabian 2003 and
Uttley et al. 2002). Consequently, their light curves were divided into
parts with $T=$1.25, 2, and 10~ksec, respectively. In order to have as
many points as possible in them, we used a bin size of 16~s for the
NGC4051 and Mrk766 light curves, and a bin size of 128~s in the case of
NGC5506. In some cases, even if $T<1/$\nb, long, well sampled light curves
were also divided into shorter parts in order to increase the number of
\rms\ estimates.

The \rms\ values were calculated for each short light curve or each
substring of a long light curve separately. The number of light curve
sub-parts ($N_{\rm p}$) used for the \bhm\ estimation of each source
is listed in column (4) of Table~1. In the case when $N_{\rm p}>1$, we
used the average \rms\ value to derive the black hole mass from Eq.
(\ref{bhm}).

\begin{figure}
\psfig{figure=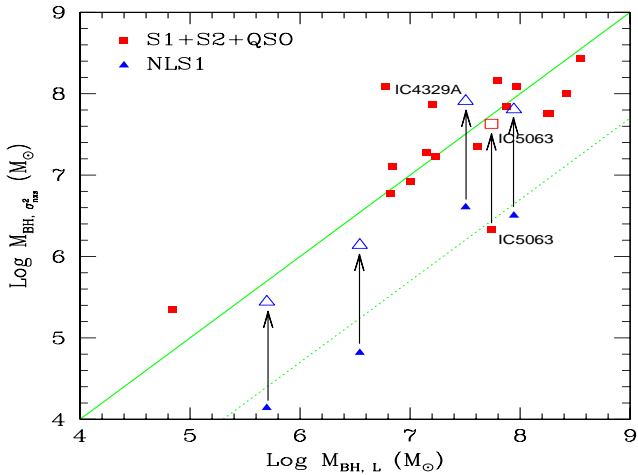,height=7.0truecm,width=8.5truecm,angle=0,%
bbllx=100pt,bblly=175pt,bburx=590pt,bbury=720pt}
\caption[]{Black hole mass estimates using the method presented in this
work (\bhms) plotted versus the literature estimates (\bhml), for the
objects listed in Table~1.  Solid squares and solid triangles show the
classical Seyfert 1/Seyfert 2/quasars and the NLS1 estimates,
respectively. Open triangles and the open square show the NLS1 and
IC5063 estimates after they have been multiplied by a factor of 20
(see text for details). The solid line shows the
\bhms = \bhml\ relation.}
\end{figure}

\subsection{Results}
\label{res}
The values of the black hole masses obtained from Eq. 
(\ref{bhm}) are given in Table~1. 

Column (1) in this table lists the names of AGN. Four objects are
classified as `Narrow Line Seyfert 1' (NLS1), while the others are
classical broad line Seyfert~1 (S1)/quasars (QSO), and Seyfert~2 (S2)
objects.  Column (2) gives the \bhm\ estimates taken mostly from Kaspi
\etal\ (2000), and they correspond to the mean value of their `mean' and
`rms' values. In cases where we found more than one \bhm\ estimate for an
object, the value listed in Table 1 corresponds to the mean of the
respective estimates. Column (3) gives our mass determination based on
X-ray variability.

Fig. 1 shows a comparison of the two mass determinations.  Generally,
the agreement between \bhms\ and \bhml\ is quite good for the
classical Seyferts (filled squares), but not in the case of NLS1s
(filled triangles). Our \bhm\ estimates for the latter are
significantly smaller than \bhml. Excluding NLS1s, we find that on
average, $\langle$\bhms/\bhml$\rangle = 2.44 \pm 1.16$. Two other
sources with the largest differences between \bhms\ and \bhml\ are
IC5063 and IC4329A; excluding also those sources from our sample we
find that on average, $\langle$\bhms/\bhml$\rangle = 1.41 \pm 0.31$.

The results for NLS1 galaxies can be reconciled with the previous
mass determination if we multiply our \bhm\ NLS1 estimates by a factor
of 20 (open triangles in Fig.~1). Interestingly, when we multiply our
IC5063 estimate by the same factor 20 as for other NLS1, the new value
becomes very similar to \bhml\ for that object.  This source is
classified as S2 with broad polarized Balmer lines (V{\' e}ron-Cetty
\& V{\' e}ron, 2001). Colina et al. (1991) found strong evidence for
the presence of a hidden, luminous source in the nucleus of
IC5063. Perhaps then, this S2 galaxy may host a NLS1 rather than a S1
nucleus, like NGC5506 (Nagar \etal\ 2002). In this light we suggest
that IC5063 is a hidden NLS1.

The source of the problems with IC4329A is more difficult to
understand. If we use the stellar velocity dispersion measurement of
225 km s$^{-1}$ (Oliva et al., 1999) for this object, together with the
\bhm-stellar velocity dispersion relation of Tremaine et al. (2002), 
we obtain \bhm $\sim 2.2 \times 10^{8}$ M$_{\odot}$, entirely
consistent with our estimate for this source. Marziani et al. (1992)
found that this source has large extinction and they gave $E(B-V)
\simeq 0.8$. After correcting the optical luminosity, $\nu L_{\nu}$,
at 5100 \AA\ for the large extinction, Nikolajuk (2004) finds \bhm =
3.9$\times 10^8$ M$_{\odot}$, using the accretion disk method. This
value is again consistent with the \bhm\ estimate listed in
Table~1. Therefore, the value given by Kaspi et al. (2000) may be too
small.

When we exclude IC4329A from our sample but we consider S1 mass
estimates, together with the scaled NLS1 and IC5063 estimates, then
$\langle$\bhms/\bhml$\rangle = 1.30 \pm 0.25$.  This result suggests
that the new method we propose for the \bhm\ estimation in AGN
provides estimates which are consistent with the estimates based on
the reverberation mapping technique or the \bhm-stellar velocity
dispersion relation.

\vspace{-1em}
\section{Discussion}

In this work we propose a new method for the estimation of black hole
mass in AGN. The method is based on the use of the `normalized excess
variance', \rms, as estimated from X-ray light (Eq.~\ref{rms}). We
find a simple expression between \rms\ and \bhm\ (Eq.~4), if we assume
that the power spectrum in AGN has a `universal' form: (1) there is a
break frequency which scales with \bhm, (2) the PSD slope above the
break frequency is $-2$, and (3) the PSD amplitude below the break
frequency is roughly constant in all objects, independent of \bhm\
(see Section 2). Note that the method is applicable to $2-10$ keV
light curves, as the PSD may be different in other energy bands
(e.g. Nandra \& Papadakis 2001; Vaughan \& Fabian 2003; Vaughan et
al. 2003).

Using our method on archival \xte, and \asca, $2-10$ keV light curves,
we estimated \bhm\ in 21 AGN, with previously known \bhm\
estimates. The comparison between the new and past \bhm\ estimates
indicates that the method works very well for `classical' AGN,
i.e. S1, S2 galaxies and quasars. The \bhm\ values that we find with
the proposed X-ray variability method are comparable to the values
determined with other methods.  However, there seems to be a
significant disagreement for NLS1 objects.  Their mass estimates,
determined from the X-ray variability method, are significantly
smaller than the \bhm\ estimates which are based on other methods. The
basic issue, that there is a shift in the X-ray variability amplitude
of NLS1 and `classical' AGN of the same mass, was already noticed by
Czerny et al. (2001). Furthermore, Papadakis (2004) and McHardy et
al. (2003) have showed that the break frequency \nb\ of NLS1 is
probably higher than the respective frequency in the power spectra of
`classical' AGN with the same black hole mass. If $B^{\rm NLS1} \sim
20 \times B^{\rm class}$ (Papadakis 2004), where $B$ is the constant
defined in Section (2), and \psdamp\ is the same for both classes of
objects, then since $C=$\psdamp$B$, we should expect that $C^{\rm
NLS1} \sim 20 \times C^{\rm class}$.  If we adopt this view, then we
obtain an excellent agreement between the \bhm\ estimates which are
based on the present X-ray variability method and the previous
estimates for all objects, both `classical' AGN and NLS1s.

The method we present here is based on the assumption of a `universal'
power spectrum shape in AGN. Consequently, it is formally equivalent
to other methods which directly use the scaling of a certain power
spectrum characteristic (like the break frequency, or a fixed PSD
value) with \bhm\ (e.g. Hayashida et al. 1998; Czerny et al. 2001;
Markowitz et al. 2003).  However, as our method requires the
estimation of \rms\ only, it should be applicable for many more cases
than methods which require the estimation of the PSD, as the latter is
much more complicated.

Present day satellites like \xmm\ and {\it CHANDRA} will soon provide
a large number of AGN light curves with a typical exposure time of
$30-50$~ksec. Due to the low internal background of their detectors,
most of these light curves should be of high signal to noise in the
$2-10$~keV band. Furthermore, because of their long orbital periods,
light curves (as long as $\sim 1.5$ days) will be evenly
sampled. Provided that significant variations are observed, they could
be used to estimate \rms, and hence \bhm\ through Eq.~(4).

The choice of the bin size, $\Delta t$, is not crucial and should
depend mainly on the average count rate. It needs to be large enough
in order to minimize the effects of the experimental Poisson noise,
but at the same time the resulting number of the points in the light
curve, $N_{\rm data}$, should also be large (at least larger than
10). 

The choice of the length of a single data string, $T$, must satisfy
the requirement $1/T >$ \nb. If \nb\ for a given object is not known in
advance, this condition should be checked \emph{a posteriori}. For
NLS1, the length should be $\sim 20$ times smaller than the
appropriate $T$ value for classical AGN, according to the results of
the present work.

Apart from observational errors, the determination of the \rms\ is
biased by intrinsic statistical error connected with the specific
shape of the power spectrum (Vaughan et al. 2003). Averaging over
several independent variance measurements performed at various epochs
reduces the error. If we have a single long light curve the error
reduction is less efficient but can be assessed through simulations.
We performed such simulations for NGC 5506 light curve (6 times longer
than 1/\nb) with many gaps, and with typical duration of a single
string of $10^4$ s. Systematic 'leak' of the power from long to
shorter timescales appeared to be negligible, and the intrinsic
statistical scatter was $\Delta \log$ \rms $ = ^{+0.20}_{-0.19}$, at
90 per cent confidence level.

Finally, we determined the constant $C$ in Eq.~(\ref{bhm}) assuming that
\bhm\ = 10 M$_{\odot}$ for Cyg X-1. However, if a higher value of 
$\sim 20$ M$_{\odot}$ is more appropriate for this object
(e.g. Zi\'o\l kowski 2004), then $C$ and the \bhm\ estimates listed in
Table~1 should be increased by a factor of $\sim 2$. Such a change
would, however, actually introduce a discrepancy between the masses
estimated from the present X-ray variability and from other
methods. Although the mass determination by the reverberation method
may also contain a systematical error of the order of a factor $2-3$
(see Krolik 2001), this possibility underlines the importance of the
accurate determination of the properties of Cyg X-1 as this object is
routinely used as a reference source.

\section{Conclusions}
\begin{itemize}
\item We propose a new method based on measurements of \rms\ 
of X-ray variability of AGN to \bhm\ estimates. The estimates of black
hole masses in AGN obtained from our method (\bhms) are consistent
with the estimates based on other techniques such as the reverberation
mapping technique or the \bhm-stellar velocity dispersion relation.
\item The constant $C$ in Eq.~(\ref{bhm}) is equal to 0.96 for 
`classical' AGN and 19.2 (i.e. 20 times larger) for NLS1
\item We suggest that IC5063, classified as S2 galaxy, 
is in fact a hidden NLS1.
\end{itemize}

\section*{Acknowledgments}
This work was partially supported by grants 2P03D 00322 and
PBZ-KBN-054/P03/2001. 
This research has made use of the TARTARUS database, which is supported
by Jane Turner and Kirpal Nandra under NASA grants NAG5-7385 and
NAG5-7067. IEP thanks CAMK for their hospitality.

\end{document}